# Return to the Kuiper Belt: launch opportunities from 2025 to 2040

Amanda M. Zangari,* Tiffany J. Finley† and S. Alan Stern‡

*Southwest Research Institute 1050 Walnut St, Suite 300, Boulder, CO 80302, USA*

and

Mark B. Tapley§

*Southwest Research Institute P.O. Drawer 28510 San Antonio, Texas 78228-0510*

**Nomenclature**

| | | |
|---|---|---|
| AU | = | astronomical unit, average Earth - Sun distance |
| $B$ | = | impact parameter |
| C3 | = | excess launch energy |
| ΔV | = | change in velocity required to alter a spacecraft's trajectory |
| KBO | = | Kuiper Belt Object |
| PC | = | plane crossing |
| $r_p$ | = | planet radius |
| $r_q$ | = | distance from a planet at periapse |
| $\mu$ | = | gravitational constant * planet mass |
| $v_\infty$ | = | arrival velocity of a spacecraft |
| $\theta$ | = | angle between incoming and outgoing velocity vectors during a swingby |

* Research Scientist, Space Science and Engineering, azangari@boulder.swri.edu

† Principal Engineer, Space Science and Engineering, AIAA Member, tfinley@boulder.swri.edu

‡ Associate Vice President-R&D, Space Science and Engineering, AIAA Member, alan@boulder.swri.edu

§ Institute Engineer, Space Science and Engineering, AIAA Senior Member, mtapley@swri.edu

**Preliminary spacecraft trajectories for 45 Kuiper Belt Objects (KBOs) and Pluto suitable for launch between 2025 and 2040 are presented. These 46 objects comprise all objects with H magnitude < 4.0 or which have received a name from the International Astronomical Union as of May 2018. Using a custom Lambert solver, trajectories are modeled after the *New Horizons* mission to Pluto-Charon, which consisted of a fast launch with a Jupiter gravity assist. In addition to searching for Earth-Jupiter-KBO trajectories, Earth-Saturn-KBO trajectories are examined, with the option to add on a flyby to either Uranus or Neptune. With a single Jupiter gravity assist, all 45 KBOs and Pluto can be reached within a 25 year maximum mission duration. A more limited number can be reached when non-Jupiter flybys are added, and the KBOs that can be reached via these alternate routes are listed. In most cases, a single Jupiter flyby is the most efficient way to get to the Kuiper Belt, but the science return from revisiting Saturn, Uranus, or Neptune may add substantial value to a mission, so alternate flybys should be considered.**

## I. Introduction

NEW Horizons' 2015 flyby of Pluto [1] showcased an active, fascinating world, with unique characteristics that were hinted about via telescopic observations of its drastic albedo features and its suite of moons. What little we do know about some of the dwarf planets of the Kuiper Belt teases that Pluto is not the only outer world with a story to tell about our Solar System. The Kuiper Belt holds dwarf planets that rotate so swiftly they could be egg-shaped objects— (136108) Haumea and (20000) Varuna [2,3]. Satellites have been found for almost all the largest objects; some objects, such as (47171) Lempo, are even known to be trinary [4]. Rings have been found around (136108) Haumea [5], which has two moons and is the parent body of the only known Kuiper Belt collisional family [6].

Small KBOs are low-density objects, largely comprised of water ice, with objects such as (55637) 2002 UX25 (diameter ~650 km) representing the largest objects with ice-like densities [7]. For larger objects, there is a transition to a higher rock fraction [7,8]. While the majority of KBOs are composed of water ice, objects larger than (50000) Quaoar are capable of retaining volatiles such as $N_2$, CO and $CH_4$ [9]. KBOs have a variety of colors ranging from gray to very red [10].

The Kuiper Belt has been divided into several regions, based on the wide variety of orbits its members occupy, and their interactions (or lack thereof) with Neptune. The Classical Belt consists of objects that are not in resonance with Neptune, and is divided into two dynamical classes: Hot and Cold. The Cold objects are limited in their orbital inclinations, and tend to have smaller eccentricities than objects considered to be "dynamically hot" [11]. *New Horizons*' extended mission target, KBO (486958) 2014 MU69, is a member of a "kernel" of KBOs within the Cold Classical Belt that have orbits clustered around a small range of semi-major axes, eccentricities and inclinations, and are thought to be remnants of the solar disk that have formed *in situ* [11,12]. In contrast, the Hot Classical Belt is thought to be populated from objects transported from the Solar System's inner disk[13]. While classical KBOs do not have resonances with Neptune, many objects do, preserving the formation history of the Kuiper Belt [14]. For example, Pluto is in the Neptune 3:2 resonance, one of several resonances KBOs can have, including the 1:1, 5:4, 4:3, 11:8, 5:3, 7:4, 9:5, 7:4, 9:5, 11:6, 2:1, 19:9, 9:4, 7:3, 12:5, 5:2, 8:3, 3:1, 7:2, 11:3, 11:2, and 27:4 resonances [14]. Other objects are scattered from Neptune, or like (90377) Sedna, are completely detached from the disk of the Solar System, with massive inclinations and eccentricities [14]. With only Pluto having been visited, most types of KBOs have not yet been explored.

Discussion of the mission concept that would eventually become the *New Horizons* mission to Pluto began shortly after Voyager 2 flew by Neptune in 1989 [15]. Even with an arrival time of over a quarter of a century after the nearly coincident equinox and perihelion of the late 1980s, a mission to Pluto was urgent because a 2015 arrival time would precede the then-predicted collapse of Pluto's atmosphere [15]. Additionally, because Pluto was moving from equinox to solstice, each passing year meant less illuminated terrain. As an added bonus, Pluto crosses its line of nodes with Earth in 2018. The visiting spacecraft was not forced to travel far out of the plane of the Solar System, enabling exploration of the object-rich Cold Classical belt, thus maximizing the probability that a suitable KBO flyby target could be found en route.

The *New Horizons* mission itself was designed around an optimal Pluto encounter, with an arrival around opposition (by contrast, the encounter of 2014 MU69, chosen for its accessibility to the spacecraft, nearly coincides with Solar conjunction, an inconvenience that will put the spacecraft out of contact with Earth for a short time a few days after the flyby). The day and time of Pluto arrival were chosen to allow for dual radio and Solar occultations of both Pluto and its largest moon, Charon [16].

To reach Pluto, *New Horizons* was launched on an Altas V 551, with an excess launch energy (C3) of 158 $km^2/s^2$ on January 19, 2006-- the fastest ever launched. Just over a year later on February 28, 2007, it flew by Jupiter at a distance of approximately 32.25 Jupiter radii (about 2.3 million km) [16]. The next 8.5 years were quiet, punctuated by annual checkouts to minimize the electronics' "on time" [17] until the arrival at Pluto on July 14, 2015, flying by at 13.78 km/s [16] and a distance of 13691 km (to body center) at closest approach [1].

*New Horizons* is the fifth spacecraft to have enough velocity to escape the Solar System, and is preceded by *Voyager 1*, *Voyager 2*, *Pioneer 10* and *Pioneer 11*.

*Voyager 1* and *Voyager 2* were launched with C3s of 105 $km^2/s^2$ and 102 $km^2/s^2$[**] [18]. Although *New Horizons* had a more powerful launch, both *Voyager 1* and *Voyager 2* currently and will continue to exceed the *New Horizons* heliocentric velocity, due to their gravity assists from Jupiter, Saturn, and in the case of *Voyager 2*, Uranus and Neptune (the Neptune flyby actually slowed the spacecraft down a bit) [18].

Preceding the *Voyager* spacecraft were *Pioneer* 10 and *Pioneer* 11, with launch C3s of 95 $km^2/s^2$ and 87 $km^2/s^2$ respectively. Pioneer 11 flew by Jupiter at a blisteringly close range of 43000 km from Jupiter's cloud tops, before buzzing Saturn at a distance of 20000 km and leaving the Solar System [19]. *Pioneer 10* and *Pioneer 11*, at 259 kg each were just over the half the weight of *New Horizons*' 400 kg dry mass plus 78 kg of propellant [19, 20]

As launches can be easily scrubbed by poor weather and adverse conditions, *New Horizons* considered several alternate methods of getting to Pluto, as well as other contingencies. If launches failed for the first two weeks of the main Jupiter gravity window, launches over the next few days could still use Jupiter, but would arrive a few years later. After the Jupiter gravitational assist window closed, direct launch to Pluto later that year and in 2007 and 2008 were still possible, with later arrival years. In 2009 and 2010, Saturn gravity assists were possible. Alternatively, *New Horizons* could have used multi-year Earth gravity assists. If none of these options were acceptable, a mission to an unspecified KBO via Uranus and Saturn was also possible. Fortunately, *New Horizons* was launched within its first choice Jupiter gravity assist window [16].

---

[**] The C3s for the *Voyager* and *Pioneer* spacecraft reported herein were calculated with MAKO using the dates (but not time of day) in the referenced papers, and are approximate.

Two *New Horizons* type mission concepts were also proposed in the 2000s. *New Horizons 2*, designed to make use of flight spares, was to fly by Jupiter, then fly by Uranus at equinox before going onto (47171) Lempo (formerly 1999 TC36), a trinary KBO system, though 2002 UX25 was also considered [5, 21]. Later, *Argo* was a proposed New Frontiers 4 mission concept. It merged flybys of Saturn, Neptune/Triton, a Trojan, and a KBO. Potential objects considered were 2001 TW148, 2005 TB190, and 2001QS322. Fast Jupiter flybys were also considered. *Argo* required a flyby velocity of < 17 km/s at Neptune, with 10 year travel time. Using the Atlas 541, potential launches had C3 < 122 $km^2/s^2$ [22].

Inspired by *New Horizons*, McGranaghan et al [23] present a Jupiter flyby study of the 5 largest KBOs other than Pluto—(1361099) Eris, (136108) Haumea, (136472) Makemake, (90377) Sedna, and (50000) Quaoar. For three Jupiter-KBO synodic alignment windows between 2014 and 2050, three viable launch years were analyzed. McGranaghan et al [23] looked at unpowered Jupiter flybys, periapse only constrained by radius of the planet. Both the C3 and the KBO arrival velocity of the search trajectories were limited to that of *New Horizons*, and chose trajectories with similar arrival velocities. An arbitrary cap of 25 years was placed on mission time of flight [23].

Since then, several follow-up studies have been conducted. Greaves el al [24] analyzed Jupiter gravity assist trajectories from 2022 to 2055 to (38628) Huya, (28978) Ixion, (90482) Orcus, (38083) Radamanthus, (120347) Salacia, (90377) Sedna, and (20000) Varuna, with an eye toward orbital capture. Kreitzman et al [25] consider (50000) Quaoar, (90377) Sedna, (136472) Makemake, (136108) Haumea, (28978) Ixion, and (38628) Huya, along with 2002

TC302, 2007 UK126, 2002 AW197, and 2003 AZ84. For a selection of the larger objects' trajectories, Kreitzman et al [25] calculate the particular radiation dose received during the path through the Jupiter system, and compare it to a common cut-off limit. Baskaran et al [26,27] compare high (C3 ~100 $km^2/^2$) and low (C3 ~50 $km^2/s^2$) thrust Jupiter gravity assist trajectories to Haumea, Huya, Ixion Pluto, Quaoar, and Varuna departing between 2022-2025, and calculate their Jupiter radiation doses. A Jupiter-Saturn gravity assist trajectory to Huya is also considered [r4].

To reduce mission cost and increase mission science return, Costigan et al [28] outline a mission in which three spacecraft sharing a launch vehicle to travel to three different KBOs. To cut down on the C3 required to, the spacecraft would undergo an Earth gravity assist, and then a Jovian swingby, before reaching 2002 UX25, 1998 WW31 and (47171) Lempo in 17.3, 25.3 and 26.3 years.

Sanchez et al [29] present a study for a mission to (136108) Haumea launching between 2020 and 2035 that includes an analysis of direct trajectories, and Jupiter or Saturn gravity assists.

Our goal is to look forward to the next generation of KBO missions. We want to show what missions are possible, how long they might take and when they could happen. While the study by McGranaghan et al [23] showcases just five of the most desirable targets, we aim to expand the list of choices, and analyze routes by all giant planets, not just Jupiter. Uranus and Neptune have not been visited since the *Voyagers*. While the *Cassini* mission is only recently ended, an opportunity to return to Saturn and revisit the system is not unwelcome.

## II. Methodology

To explore KBO trajectories, we used MAKO [30]. This tool was designed for use at Southwest Research Institute with future KBO missions in mind, though it can be generalized to other applications. MAKO is Python code, making extensive use of the Spiceypy [31] wrapper for the SPICE (Spacecraft, Planet, Instrument, Camera-matrix, Events) system [32].

### A. MAKO procedure

Briefly, MAKO can quickly identify trajectories using the patched conic method [33]. MAKO's Lambert solver identifies the departure and arrival velocities between two planets. By treating Earth launch and time of flight (TOF) to Jupiter as independent variables, we can solve for the arrival date at a third object that minimizes the ΔV required between the swingby planet arrival and departure. If this third object is a giant planet, such as Saturn, Uranus, or Neptune, a fourth object can be added on top of a three-body mission.

A valid trajectory solution must meet several criteria, some of them physical, some set by the user. At launch, the C3, or excess energy (here the square of the Earth departure velocity as calculated by MAKO), must be within the range of a specified launch vehicle's performance. The declination of launch asymptote, or J2000 declination of the Earth departure velocity vector must be less than the latitude of the launch site.

MAKO seeks to find the arrival time that minimizes the *magnitude* of the velocity difference between the swingby arrival and swingby departure. The direction change is handled by the planet. Knowing the angle between the departure and arrival velocity vectors, we can solve for the closest approach distance of the swingby, $r_q$:

$$r_q = \frac{\mu}{v_\infty^2}\left(\frac{1}{sin\frac{\theta}{2}} - 1\right) \tag{1}$$

where $\mu$ is the universal gravitational constant times the planet mass, $v_\infty$ is the arrival velocity, and $\theta$ is the angle between the incoming and outgoing velocity vectors. Closer flybys are required for larger turn angles and higher velocities. The swingby closest-approach radius must be greater than $r_p$, the radius of the swingby planet.

In addition to this forced minimum, there may be reasons to add additional constraints beyond what is simply physically allowed. The intense radiation environment of Jupiter may be a concern, and so an absolute minimum might be set, to avoid needing a radiation vault such as that on the Juno spacecraft [34]. In contrast, McGranaghan et al [23] do not restrict closest approach in any way, other than requiring it to be larger than the planet radius.

In addition to constraining closest approach, spacecraft must avoid flying through planetary rings, though safe passage may be found between certain rings. Geometrically speaking, the ring plane crossing radius does not coincide with closest approach, so this value must be calculated separately. The mathematical process MAKO uses to calculate the ring plane crossing is detailed in the tool description paper [30]. MAKO allows the user to set a simple ring plane crossing minimum distance.

By solving for the minimum ΔV, we reduce it to essentially zero (fractions of meters per second). However, if the mission elapsed time producing minimum ΔV is outside the TOF set by the user, the TOF will be pegged at one of the bounds, and the ΔV will have a substantial value. Some ΔV, e.g. a few tens of m/s, is not unreasonable for a spacecraft to carry, but ideally there should be no ΔV needed for initial design.

**B. Parameter Selection**

Our launch dates and Earth-Jupiter/Saturn TOF make up the independent variables.

The Earth-Jupiter synodic period is approximately 11.86 years, thus Jupiter launch opportunities for each KBO repeat for some KBOs during this time frame. We have chosen to look for all launches between 2025-01-01 and 2040-12-31. A mission competed at publication time could plausibly be built by 2025-- it took exactly five years between the announcement of opportunity and *New Horizons*' launch [20]. Extending the time frame to the end of the 2030s allows for opportunities for the objects with the earliest launches to repeat themselves. The window is also large enough to include Jupiter-Saturn, Jupiter-Uranus, and Jupiter-Neptune launch windows.

For Earth-Jupiter TOF, we have chosen travel times between 340 and 595 days inclusive, to represent a fast, direct flight time to Jupiter. For Saturn, we explored travel times from 700 to 1795 days inclusive.

We tested every 5 days in this range of departure dates and Earth-Jupiter TOF, which allowed good coverage of each Jupiter launch window and kept run times reasonable.

We allowed all C3s between 75 and 200 $km^2/s^2$. *New Horizons*, the fastest spacecraft ever launched, had a C3 of 158 $km^2/s^2$. We wanted to showcase the possibilities available to lighter spacecraft and more powerful launch vehicles, and brought the upper limit up to 200 $km^2/s^2$. Below a C3 of 75 $km^2/s^2$, we are no longer analyzing a fast KBO mission. The 595-day Earth-Jupiter TOF limit cuts off before the lower C3 limit is reached.

We set the DLA limit to +/-28.5°, the approximate latitude of Kennedy Space Center.

For TOFs of mission legs beyond Jupiter, we consider 0.2 to 25 years between targets. The lower extreme is to accommodate fast trajectories between Jupiter and Saturn. The upper extreme is designed to allow for inclusion of the farthest objects, such as Eris. After trajectories are found, we exclude any mission with a total duration greater than 25 years. Duration is defined to be time span between Earth launch and arrival at the final target object, though mission operations do not end immediately upon flyby.

For ΔV, we set an upper limit of 2 m/s. This limit is well above the convergence point of the TOF solver for the last body. Our TOF range is generous enough that we do not need to worry about pegged dates.

We limit Jupiter closest approaches to 6 Jupiter radii as a first-order guideline against radiation exposure. Our limit is just outside both the orbit of Io (5.91 Jupiter radii) and the peak S+ region of the Io plasma torus (5.71 Jupiter radii) [35]. A more-detailed study that calculates the expected radiation exposure for a particular trajectory, along the lines of Kreitzman et al [25], Baskaran et al [26,27] or Costigan [28] is beyond the scope of this work. We did not have any radiation information or warnings for Saturn, Uranus or Neptune, so we set the closest approach limits at an arbitrary 1.1 radii for each of those planets.

Passage through Saturn's rings can only be performed at certain locations: inside all rings and near the cloud-tops as executed in the Cassini Grand Finale, between the F and G rings, and from the start of the sparse E ring (roughly 3 Saturn radii) and beyond[††]. To simplify searches, we chose to only entertain trajectories that crossed Saturn's ring plane beyond the E ring start at 180,000 km from the planet's center [36].

While Uranus's rings were recently discovered to extend out to 103,000 km [30], the most dangerous areas are inside the ε ring (located at 51,149 km or about 2 Uranus radii from planet center [35]). It is uncertain whether the recently discovered, low optical-depth rings (67,300-69,900 km and 86,000-103,00 km from body center [37]) pose a risk. However, spaces between the ε, ν and μ ring should be safe for spacecraft passage[‡‡]. In its current state, MAKO allows for a single outer ring limit only, so we have set the Uranus ring plane crossing limit at 51,250 km (adding 100 km for the ring width [38]), but manually removed trajectories that fall between the boundaries of the ν and μ rings (2.6-2.7 and 3.4-4.0 Uranus radii).

[††] J. Spencer, personal communication, May 2018

The rings of Jupiter extend only out to 226,000 km [36], well below the previously-imposed closest approach limit, so Jupiter's rings do not limit any flyby opportunities. For Neptune, we set the plane crossing minimum distance at the outer edge of the ring system: 62,930 km [36].

### C. Target Selection

The Minor Planet Center's list of distant objects[§§] was accessed on May 3, 2018. From this list, all objects with $H < 4$, $q > 28.5$ AU, were selected, as well as all named objects with $q > 28.5$ AU. The H-magnitude criterion allowed for the selection of the largest KBOs, while the name criterion was a quick heuristic to identify smaller, but interesting multi-object systems that have been highly studied, such as (47171) Lempo, (385446) Manwe, (120347) Salacia, (79360) Sila-Nunam. (15760) Albion, formerly 1992 QB1 was also included through this selection process. Including Pluto, we have a list of 46 objects to study.

## III. Results

Using the parameters and methodology described in Section II, we present the results of several MAKO runs for potential KBO missions. We have explored seven different scenarios that cover different types of swingbys: A. Jupiter-KBO, B. Jupiter-Saturn-KBO, C. Jupiter-

‡‡ M. Showalter, personal communication, September 2018

§§ https://www.minorplanetcenter.net/iau/MPCORB/Distant.txt

Uranus-KBO, D. Jupiter-Neptune-KBO, E. Saturn-KBO, F. Saturn-Uranus-KBO, G. Saturn-Neptune-KBO. Information is presented in aggregate for each object, with the fastest launch time per bin of C3. At the end of this section, we present a selection of promising missions in greater detail.

### A. Jupiter Mission Summaries

Launching a mission to any KBO with a Jupiter gravity assist is a simple matter: once every synodic period (just under 400 days), a new Jupiter launch window opens up. As Jupiter travels around the sun, it comes within range of every KBO on our list for a period of a few years throughout its 12-year orbit. Each KBO thus has an availability window for about 2-5 years' worth of Earth-Jupiter launches. Every KBO we looked at is accessible by Jupiter gravity assist at some point between 2025-2040 in under 25 years with a C3 of < 165 $km^2/s^2$. With launches and arrivals in five-day increments, we found 70 trajectories for the furthest KBOs (most strongly affected by the 25 year maximum mission cut-off). At the other extreme, relatively near KBOs that were within launch range in the 2020s experienced another Jupiter gravity assist window within our 15-year search time frame, and over 1000 trajectories were found.

Given the number of possible trajectories for each object, we must present a summary of the findings for each event. Table 1 lists every object searched, the range of possible launch years and the fastest travel time for four categories of rockets. The first category (C3 < 200 $km^2/s^2$) looks at all viable trajectories and shows the fastest trajectory to each object. "Min duration", used throughout the tables simply refers to the travel time from Earth to the object listed in the first column, and does not include time for downlink or other post-fly activities, such as calibration campaigns, Earth lookback outreach images, or distant KBO observations, which may continue to take place 1-2 years or more after flyby. The second category (C3 < 165 $km^2/s^2$) shows the fastest trajectory assuming a launch vehicle that is about as powerful as *New Horizons*. The third and fourth categories (C3 < 140 $km^2/s^2$, C3 < 120 $km^2/s^2$) show what can be done with lower performance launch vehicles. At the most optimal launch geometry, higher launch energies correlate with lower travel times, but a well-aligned, lower C3 trajectory can perform almost as well. In some cases, the fastest trajectory required excess energies of less than 120 $km^2/s^2$, and thus the same value was included in all previous columns.

Because the search window spans more than a Jupiter synodic period, many objects are accessible during two separate launch seasons, separated by an Earth year or more in which a launch is not possible. In these cases, the first season is denoted by "I", the second by "II". Often one of these periods has an inferior range of options, owing to the truncation of part of the season by the search time constraints, unfavorable geometry or larger heliocentric range caused by an eccentric orbit.

Assuming optimal launch geometry, it takes the longest to reach the farthest KBOs. With current rocket capabilities, it takes a minimum of > 20 years to get to (255088) 2007 OR10, the most distant object in our survey. On the other hand, Neptune Trojan (385571) Otrera can be reached in under six years.

It is worth noting that options are often poor in the last year of a set of launch windows, which explains why Pluto missions with much lower C3s and shorter travel times than *New Horizons* had (158 $km^2/s^2$, 9.5 years, ~33 AU) are still feasible even though Pluto has retreated further from the sun, and will be higher out of the ecliptic since the encounter. *New Horizons* was launched in the last year that a Jupiter gravity assist was available [16].

**Table 1: List of KBOs with Jupiter swingbys, their possible launch years, and the minimum mission duration for various amounts of maximum C3.**

| **Object** | **Launch Range** | **Min Duration C3 < 200 km²/s² (years)** | **Min Duration C3 < 165 km²/s² (years)** | **Min Duration C3 < 140 km²/s² (years)** | **Min Duration C3 < 120 km²/s² (years)** | **Final Target Sun range (AU)** |
|---|---|---|---|---|---|---|
| Jupiter | 2025 - 2040 | 0.9 | 1 | 1.1 | 1.2 | 4.9 - 5.5 |
| 2002 AW197 I | 2025 - 2026 | 10.3 | 13.4 | 21.1 | - | 42.1 - 43.5 |
| 2002 AW197 II | 2035 - 2039 | 9.2 | 11 | 12.4 | 13.2 | 41.4 - 42.5 |
| 2002 MS4 I | 2027 - 2031 | 9.1 | 10.9 | 12.1 | 12.8 | 42.5 - 44.7 |
| 2002 MS4 II | 2040 | 13 | 13 | 13 | 13 | 42.0 - 42.8 |
| 2002 TC302 | 2032 - 2036 | 9.2 | 11.3 | 12.3 | 12.9 | 39.1 - 40.0 |
| 2002 TX300 | 2032 - 2035 | 11.6 | 11.6 | 12 | 15.8 | 45.5 - 46.9 |
| 2002 UX25 | 2032 - 2036 | 7.9 | 9.4 | 10.4 | 10.4 | 36.7 - 37.7 |
| 2003 AZ84 I | 2025 | 14.6 | 23.5 | - | - | 39.8 - 41.5 |
| 2003 AZ84 II | 2035 - 2038 | 8.2 | 9.7 | 11.6 | 11.6 | 37.4 - 40.7 |
| 2003 OP32 | 2031 - 2033 | 9.2 | 11.1 | 11.1 | 11.4 | 44.6 - 46.0 |
| 2005 QU182 | 2032 - 2033 | 14.7 | 17.1 | 17.1 | 17.9 | 68.4 - 73.8 |
| 2005 RN43 | 2031 - 2033 | 7.8 | 9 | 10.8 | 11.1 | 40.6 - 40.8 |
| 2005 UQ513 | 2031 - 2034 | 10.4 | 10.8 | 13.5 | 14.2 | 43.3 - 45.4 |
| 2007 OR10 | 2030 - 2031 | 18.1 | 22.2 | 24.8 | - | 94.1 - 95.2 |
| 2007 UK126 | 2034 - 2037 | 7.8 | 9.5 | 9.7 | 9.8 | 37.5 - 39.3 |
| 2010 EK139 I | 2026 - 2031 | 7.4 | 8.9 | 9.2 | 11.5 | 32.5 - 35.2 |
| 2010 EK139 II | 2039 - 2040 | 8.8 | 8.8 | 8.8 | 10.2 | 33.7 - 37.3 |
| 2010 RF43 | 2030 - 2032 | 12.5 | 14.2 | 14.2 | 15.8 | 57.9 - 59.4 |
| 2013 FY27 I | 2025 | 17.9 | 24.1 | - | - | 75.8 - 76.9 |
| 2013 FY27 II | 2035 - 2037 | 17.2 | 17.4 | 18.4 | 23.4 | 73.6 - 75.2 |
| 2014 EZ51 I | 2026 - 2030 | 11.2 | 11.4 | 13.6 | 15.3 | 48.6 - 51.8 |
| 2014 EZ51 II | 2038 - 2040 | 10.9 | 11.6 | 12.1 | 15.1 | 46.4 - 49.2 |
| 2014 UZ224 | 2032 - 2034 | 17.1 | 18 | 23 | 23 | 71.2 - 75.5 |
| 2015 KH162 I | 2026 - 2029 | 14.6 | 16.6 | 20.1 | 23 | 66.0 - 68.6 |
| 2015 KH162 II | 2038 - 2040 | 16.1 | 17 | 22.2 | 22.5 | 69.1 - 71.2 |
| 2015 RR245 | 2031 - 2034 | 11.2 | 11.2 | 11.7 | 14.2 | 43.2 - 50.7 |
| Albion | 2032 - 2035 | 9.7 | 10 | 10 | 11.5 | 42.6 - 43.7 |
| Altjira | 2033 - 2036 | 10.4 | 10.4 | 11.1 | 14.1 | 46.8 - 47.0 |
| Arawn | 2028 - 2032 | 8.2 | 8.5 | 8.9 | 11 | 37.6 - 39.8 |
| Borasisi | 2031 - 2033 | 9.1 | 9.2 | 10.7 | 13.1 | 44.0 - 45.3 |
| Chaos | 2034 - 2037 | 8.2 | 9.9 | 11.5 | 11.5 | 41.1 - 41.9 |
| Deucalion I | 2026 - 2030 | 8.7 | 9.3 | 11.1 | 12.1 | 41.4 - 41.9 |
| Deucalion II | 2038 - 2040 | 8.9 | 9.9 | 9.9 | 12.1 | 41.3 - 41.5 |
| Eris | 2032 - 2033 | 20.3 | 20.4 | 21.8 | - | 91.8 - 92.6 |
| Haumea I | 2025 - 2029 | 11.6 | 12.1 | 15.5 | 16.3 | 46.2 - 48.4 |
| Haumea II | 2037 - 2040 | 10.9 | 12 | 12.8 | 16.8 | 44.1 - 46.7 |
| Huya I | 2027 - 2032 | 6.7 | 7.5 | 7.8 | 9.5 | 31.3 - 36.9 |
| Huya II | 2040 | 10.5 | 10.5 | 10.5 | 10.5 | 35.3 - 36.0 |
| Ixion I | 2027 - 2032 | 7.3 | 8.6 | 8.6 | 10.1 | 30.8 - 34.8 |
| Ixion II | 2040 | 10.2 | 10.2 | 10.2 | 10.2 | 31.4 - 31.8 |
| Lempo | 2033 - 2036 | 6.8 | 7.2 | 8.5 | 10.2 | 33.4 - 37.9 |
| Logos I | 2025 - 2027 | 9.8 | 10.8 | 11.2 | 14.7 | 45.3 - 47.0 |
| Logos II | 2037 - 2040 | 9.2 | 11.1 | 12.4 | 13.7 | 46.5 - 48.1 |
| Makemake I | 2025 - 2027 | 12.8 | 13.3 | 15.2 | 23.6 | 52.4 - 52.8 |
| Makemake II | 2036 - 2039 | 11.6 | 13.6 | 13.7 | 19.2 | 51.7 - 52.5 |
| Manwe | 2031 - 2034 | 8.9 | 8.9 | 9.6 | 11.9 | 39.6 - 41.1 |
| Mors-Somnus I | 2027 - 2031 | 9.1 | 10.9 | 12.1 | 12.8 | 42.5 - 44.7 |
| Mors-Somnus II | 2040 | 13 | 13 | 13 | 13 | 42.0 - 42.8 |
| Orcus I | 2025 - 2026 | 10.8 | 13.7 | 20.6 | - | 46.1 - 47.5 |
| Orcus II | 2035 - 2039 | 9.8 | 11.7 | 13.3 | 13.8 | 44.3 - 46.6 |
| Otrera | 2033 - 2036 | 5.8 | 6.5 | 7.7 | 8.7 | 29.6 - 30.2 |
| Pluto | 2028 - 2032 | 8.4 | 9 | 9.2 | 11.5 | 38.8 - 43.1 |
| Praamzius I | 2025 - 2026 | 11.5 | 16.3 | - | - | 43 |
| Praamzius II | 2035 - 2038 | 9.6 | 9.6 | 10.6 | 13.1 | 42.9 - 43.0 |
| Quaoar I | 2027 - 2031 | 8.8 | 10.3 | 10.3 | 12.1 | 41.9 - 42.3 |
| Quaoar II | 2040 | 11.6 | 11.6 | 11.6 | 11.6 | 41.8 - 41.9 |
| Rhadamanthus I | 2025 - 2029 | 9.7 | 9.9 | 11.4 | 14.1 | 43.3 - 44.6 |
| Rhadamanthus II | 2037 - 2040 | 9.2 | 11.1 | 11.4 | 13 | 44.3 - 44.9 |
| Salacia | 2031 - 2034 | 11.4 | 11.6 | 11.7 | 14.6 | 46.2 - 46.5 |
| Sedna | 2033 - 2034 | 17.3 | 19.4 | 19.5 | 20.7 | 77.0 - 77.9 |
| Sila-Nunam I | 2025 - 2026 | 10.9 | 14.9 | - | - | 43.4 |
| Sila-Nunam II | 2035 - 2038 | 9.5 | 10.2 | 10.4 | 12.6 | 43.4 |
| Teharonhiawako | 2031 - 2033 | 8.9 | 10.4 | 11.9 | 11.9 | 45.1 - 45.2 |
| Varda I | 2027 - 2031 | 10.1 | 10.2 | 12.3 | 13.6 | 41.8 - 44.0 |
| Varda II | 2039 - 2040 | 10.9 | 10.9 | 10.9 | 12.7 | 40.8 - 42.3 |
| Varuna I | 2025 | 16.7 | - | - | - | 44.8 - 45.0 |
| Varuna II | 2035 - 2038 | 9.2 | 11.2 | 12.3 | 12.3 | 44.9 - 45.2 |

**B. Jupiter-Saturn Mission Summaries**

While Saturn is technically accessible from Jupiter between 2025-2035 (I), and 2037-2040 (II), launches in the latter window provide the best opportunities to visit a KBO with dual Jupiter and Saturn swingbys. Of the 45 KBOs analyzed, 19 KBOs are reachable from Saturn for launches by 2040. Table 2, repeating the format of Table 1, lists these targets and minimum mission durations for the four classes of launch vehicles.

With our 25 year limit (chosen to be a catch-all time limit to accommodate scattered disk objects), we see that some trajectories are included are physically feasible, but not practical. Skipping Saturn saves over 10 years travel time to (174567) Varda, (208996) 2003 AZ84 and several other KBOs are at a disadvantage compared to their Table 1 travel times. Conversely, Saturn gravity assists provide a time advantage to objects such as (136108) Haumea, (136472) Makemake and (38083) Rhadamanthus for less-powerful launch vehicles.

**Table 2: List of KBOs with Jupiter and Saturn swingbys, their possible launch years, and the minimum mission duration for various amounts of maximum C3.**

| **Object** | **Launch Range** | **Min Duration C3 < 200 km$^2$/s$^2$ (years)** | **Min Duration C3 < 165 km$^2$/s$^2$ (years)** | **Min Duration C3 < 140 km$^2$/s$^2$ (years)** | **Min Duration C3 < 120 km$^2$/s$^2$ (years)** | Final Target Sun range (AU) |
|---|---|---|---|---|---|---|
| Saturn I | 2025 - 2035 | 8.4 | 10.8 | 11 | 11 | 9.0 - 10.0 |
| Saturn II | 2037 - 2040 | 2.2 | 2.4 | 2.5 | 2.8 | 9.6 - 10.0 |
| 2002 AW197 | 2037 - 2039 | 17.5 | 17.5 | 17.5 | 18.1 | 41.4 - 41.9 |
| 2002 MS4 | 2038 - 2040 | 18.6 | 18.6 | 19 | 19.5 | 41.2 - 42.2 |
| 2003 AZ84 | 2037 - 2038 | 23.8 | 23.9 | 23.9 | - | 37.4 - 37.7 |
| 2010 EK139 I | 2026 - 2027 | 21.6 | 24.5 | - | - | 33.6 - 34.1 |
| 2010 EK139 II | 2039 - 2040 | 18.6 | 18.6 | 18.6 | 18.7 | 36.0 - 38.7 |
| 2014 EZ51 | 2037 - 2040 | 13 | 13 | 13 | 13.9 | 46.4 - 49.3 |
| 2015 KH162 | 2037 - 2039 | 17 | 17 | 17.2 | 17.8 | 68.9 - 71.0 |
| Deucalion I | 2025 | 22.7 | 24.2 | - | - | 41.5 |
| Deucalion II | 2037 - 2040 | 11.7 | 11.7 | 11.8 | 11.9 | 41.3 - 41.5 |
| Haumea | 2037 - 2040 | 12.4 | 12.5 | 12.6 | 13 | 44.2 - 46.7 |
| Huya | 2038 - 2039 | 19.4 | 20 | 20 | 20 | 37.2 - 37.9 |
| Ixion I | 2027 - 2029 | 21.9 | 24.4 | - | - | 31.5 - 31.8 |
| Ixion II | 2035 - 2040 | 21.8 | 21.8 | 21.8 | 21.8 | 30.0 - 30.5 |
| Logos | 2037 - 2039 | 13.4 | 13.4 | 13.4 | 13.8 | 46.8 - 48.1 |
| Makemake | 2037 - 2039 | 16.8 | 16.8 | 16.8 | 17.3 | 51.7 - 52.3 |
| Mors-Somnus | 2038 - 2040 | 18.6 | 18.6 | 19 | 19.5 | 41.2 - 42.2 |
| Orcus | 2037 - 2039 | 19.2 | 19.2 | 19.2 | 19.9 | 44.3 - 45.5 |
| Praamzius | 2037 - 2038 | 22.3 | 22.3 | 22.3 | 23.3 | 42.9 - 43.0 |
| Quaoar | 2038 - 2040 | 18.3 | 18.4 | 18.5 | 23.8 | 41.8 - 41.9 |
| Rhadamanthus | 2037 - 2040 | 11 | 11 | 11.1 | 11.6 | 44.3 - 44.9 |
| Sila-Nunam | 2037 - 2038 | 21 | 21 | 21 | 21.9 | 43.4 |
| Varda | 2037 - 2040 | 14.5 | 14.6 | 14.6 | 14.9 | 40.8 - 42.1 |

### C. Jupiter-Uranus Mission Summaries

The fourteen KBOs reachable via a possible swingby of Uranus are listed in Table 3, with the same columns as Table 1. The fastest trajectories take between 3.8 and 4.7 years to arrive at Uranus before arrival at a KBO. While Jupiter-Uranus missions are possible from 2025-2026 (“Uranus I”), no KBOs are accessible then. Uranus is again accessible from Jupiter with launches from 2034-2040 (“Uranus II”), however, the best KBO mission opportunities with Uranus flybys have launch opportunities between 2034-2036.

As with the Jupiter-Saturn missions, some objects, such as (148780) Altjira, (420356) Praamzius, (79360) Sila-Nunam and (47171) Lempo, are physically feasible, but not practical, their inclusion on this list is due to broad constraints on total mission duration. While Uranus encounters can be paired with interesting objects, such as (20000) Varuna, a Uranus flyby will likely add time to a mission.

As many KBOs have highly inclined orbits relative to the plane of the ecliptic, the spacecraft must rely on the final giant planet flyby to bend a trajectory that started out as nearly co-planar to the ecliptic into what may be a more highly-inclined one. For Uranus, a large number of trajectories were eliminated due to the ring plane crossing constraint, which was not the case with Saturn and Neptune. To achieve the same turn angle as a forbidden close encounter, a slower Uranus approach velocity is needed at the larger radii needed to avoid the ring hazard. Thus Uranus encounters require longer travel times to a KBO, in spite of dual gravitational assists. Permitting a swingby between the ν and μ rings could reduce travel time by several years. While we can approximate the location of a ring plane crossing, and make a crude cut, the feasibility and TOF benefits of flying between the rings requires future study.

**Table 3: List of KBOs with Jupiter-Uranus swingbys, their possible launch years, and the minimum mission duration for various amounts of maximum C3.**

| Object | Launch Range | Min Duration C3 < 200 $km^2/s^2$ (years) | Min Duration C3 < 165 $km^2/s^2$ (years) | Min Duration C3 < 140 $km^2/s^2$ (years) | Min Duration C3 < 120 $km^2/s^2$ (years) | Final Target Sun range (AU) |
|---|---|---|---|---|---|---|
| Uranus I | 2025 - 2026 | 15.6 | 22 | - | - | 18.3 - 18.5 |
| Uranus II | 2034 - 2040 | 3.8 | 4.1 | 4.7 | 5.7 | 18.3 - 18.8 |
| 2002 AW197 | 2035 - 2037 | 22.6 | 22.6 | 22.6 | 22.6 | 41.5 - 41.7 |
| 2002 TC302 | 2036 | 24.1 | 24.3 | - | - | 39.1 |
| 2002 UX25 | 2035 - 2036 | 21.4 | 22.1 | 23.4 | 23.4 | 36.7 |
| 2003 AZ84 | 2034 - 2038 | 13.1 | 13.1 | 13.1 | 13.2 | 37.5 - 40.2 |
| 2007 UK126 | 2034 - 2037 | 9.6 | 9.7 | 10.6 | 12.7 | 37.5 - 39.2 |
| 2013 FY27 | 2034 – 2036 | 22 | 22 | 22 | 22.2 | 73.8 – 74.7 |
| Altjira | 2034 - 2036 | 16.1 | 18 | 18 | 19.4 | 46.9 - 47.0 |
| Chaos | 2034 - 2037 | 9.5 | 9.6 | 11.3 | 13.5 | 41.1 - 41.9 |
| Lempo | 2034 - 2036 | 15.3 | 17.2 | 17.2 | 17.2 | 35.4 - 37.8 |
| Orcus | 2035 – 2038 | 22 | 22 | 22 | 22 | 44.5 - 45.4 |
| Otrera | 2034 - 2037 | 14.7 | 15.6 | 16.3 | 16.5 | 29.9 - 30.2 |
| Praamzius | 2035 - 2036 | 15.2 | 15.3 | 15.3 | 15.3 | 42.9 - 43.0 |
| Sila-Nunam | 2035 - 2036 | 16.7 | 16.7 | 16.7 | 16.8 | 43.4 |
| Varuna | 2034 - 2037 | 12.4 | 12.4 | 12.4 | 13.4 | 45.0 - 45.2 |

### D. Jupiter-Neptune Mission Summaries

Of our 45 KBOs and Pluto, 13 objects are reachable after an encounter with Neptune. As we can see by comparing Table 4 to its counterpart Table 1, not every mission that includes a Neptune encounter has an advantage compared to a Jupiter-only mission. However, the total TOF increase by adding on Neptune is not nearly so as severe as with Uranus. For an object like Eris, two years and a Neptune visit may provide added value to what may otherwise be considered too long, and thus too high-risk for selection. Objects like Manwe, Borasisi and 2015 RR245 have travel times closer to their Jupiter-only counterparts.

The *Argo* mission concept required a Neptune flyby velocity of less than 17 km/s at Neptune. Typical velocities for the fastest times of flight that we found were in the 19-20 km/s range. Requiring a reduced flyby speed would add around two years to the fastest mission times shown in Table 4. It would also remove some objects from consideration: no Eris mission concept that meets our current criteria meets the *Argo* Neptune arrival velocity limit.

**Table 4: List of KBOs with Jupiter and Neptune swingbys, their possible launch years, and the minimum mission duration for various amounts of maximum C3.**

| Object | Launch years | Min Duration C3 < 200 $km^2/s^2$ (years) | Min Duration C3 < 165 $km^2/s^2$ (years) | Min Duration C3 < 140 $km^2/s^2$ (years) | Min Duration C3 < 120 $km^2/s^2$ (years) | Final Target Sun range (AU) |
|---|---|---|---|---|---|---|
| Neptune | 2031 - 2035 | 6.6 | 7 | 7 | 7.8 | 29.8 - 29.9 |
| 2003 OP32 | 2032 - 2033 | 21.2 | 21.5 | 22.7 | 22.7 | 45.8 - 46.0 |
| 2005 QU182 | 2032 - 2033 | 15.7 | 15.9 | 15.9 | 18.4 | 68.5 - 73.8 |
| 2005 RN43 | 2031 - 2034 | 15.7 | 15.9 | 16.9 | 17 | 40.7 - 40.8 |
| 2005 UQ513 | 2032 - 2034 | 23.4 | 23.4 | 23.6 | 24.3 | 43.3 - 43.8 |
| 2014 UZ224 | 2032 - 2033 | 19.4 | 19.7 | 19.7 | 20.3 | 71.7 - 74.8 |
| 2015 RR245 | 2031 - 2034 | 11.6 | 12 | 12 | 13.5 | 43.0 - 50.3 |
| Albion | 2032-2034 | 21.6 | 21.6 | 21.8 | 21.9 | 43.3 - 43.6 |
| Borasisi | 2031 - 2034 | 10.7 | 11 | 11 | 12.6 | 44.2 - 45.4 |
| Eris | 2032 - 2033 | 22.2 | 22.3 | 22.3 | - | 91.7 - 92.3 |
| Manwe | 2031 - 2034 | 9.6 | 10 | 10 | 11.3 | 39.6 - 41.0 |
| Salacia | 2031 - 2033 | 16.8 | 16.8 | 17.7 | 17.9 | 46.4 - 46.5 |
| Sedna | 2032 | 24.9 | 24.9 | 24.9 | - | 77.2 |
| Teharonhiawako | 2031 - 2033 | 16.2 | 16.3 | 17.4 | 17.5 | 45.1 - 45.2 |

## Saturn Mission Summaries

Saturn's 29-year orbit comprises roughly one half of our 15-year search span. Forty-one KBOs are accessible from Saturn. While the travel time search ranged from 700 to 1795 days, the minimum travel time to Saturn with a C3 of 200 $km^2/s^2$ is 710 days. At 1795 days, the minimum C3 was just above 107 $km^2/s^2$.

Saturn does not possess the power of Jupiter for its gravity assists, so travel times via Saturn are often slower than their Jupiter counterparts. Saturn swingby launches are rarely possible with less-capable launch vehicles, and when they are, the travel times are excessively long. On the other hand, dropping Jupiter in the Jupiter-Saturn-KBO launch paradigm can decrease travel time in certain scenarios.

Table 5 lists launch years and travel times to KBOs via a Saturn assist only.

Saturn does provide the advantage of being present for a gravitational assist for different launch years than Jupiter. The objects that have viable Jupiter-Saturn launches, by necessity have some launch year overlap with their Saturn-only trajectories.

Launches to Saturn are possible from 2025-2034 (Saturn I), and from 2036-2040 (Saturn II). All 2035 Earth-Saturn launch trajectories were eliminated for having a DLA greater than 28.5°. Thus, it is not possible to launch directly to Saturn from Kennedy Space Center in 2035.

During the second Saturn II launch set, in many cases, but not all, travel times from Saturn swingbys are slower. An Earth gravity assist as demonstrated by the *Argo* mission concept may allow for a less-capable launch vehicle or cut down travel time relative to the power supplied, but the analysis of these trajectory types are beyond the scope of this paper.

**Table 5: List of KBOs with Saturn swingbys, their possible launch years, and the minimum mission duration for various amounts of maximum C3.**

| Object | Launch years | Min Duration C3 < 200 $km^2/s^2$ (years) | Min Duration C3 < 165 $km^2/s^2$ (years) | Min Duration C3 < 140 $km^2/s^2$ (years) | Min Duration C3 < 120 $km^2/s^2$ (years) | Final Target Sun range (AU) |
|---|---|---|---|---|---|---|
| Saturn I | 2025 - 2034 | 1.9 | 2.2 | 2.6 | 3.3 | 4.9 - 5.5 |
| Saturn II | 2036 - 2040 | 2 | 2.4 | 3 | - | 5.0 - 5.4 |
| 2002 AW197 I | 2026 - 2034 | 10 | 11.7 | 14.1 | 19.5 | 41.7 - 42.9 |
| 2002 AW197 II | 2036 - 2040 | 12.5 | 21 | - | - | 41.3 - 42.3 |
| 2002 MS4 | 2038 - 2040 | 16.7 | 16.7 | 16.7 | - | 41.1 - 42.3 |
| 2002 TC302 | 2025 - 2033 | 10.3 | 12.3 | 17.8 | - | 39.1 - 41.1 |
| 2002 TX300 | 2025 - 2030 | 11.2 | 14.7 | 24.5 | - | 44.7 - 46.5 |
| 2002 UX25 | 2025 - 2033 | 9.2 | 10.6 | 14.6 | - | 36.7 - 38.4 |
| 2003 AZ84 I | 2025 - 2034 | 9.2 | 10.8 | 12.9 | 18.2 | 38.3 - 41.6 |
| 2003 AZ84 II | 2036 - 2039 | 16.1 | - | - | - | 37.2 - 39.3 |
| 2003 OP32 | 2025 | 20.7 | - | - | - | 45.0 - 45.4 |
| 2005 QU182 | 2025 - 2026 | 19.7 | - | - | - | 66.9 - 70.2 |
| 2005 RN43 | 2025 - 2027 | 16.3 | - | - | - | 40.6 - 40.7 |
| 2005 UQ513 | 2025 - 2029 | 12.5 | 16.9 | - | - | 43.9 - 46.1 |
| 2007 UK126 | 2025 - 2034 | 8.8 | 10.5 | 12.5 | - | 37.5 - 38.6 |
| 2010 EK139 | 2038 - 2040 | 15.6 | 15.6 | 15.6 | - | 35.4 - 38.7 |
| 2013 FY27 | 2029 - 2034 | 16.3 | 18.7 | 21.9 | - | 74.4 - 76.2 |
| 2014 EZ51 | 2037 - 2040 | 11.6 | 13.4 | 17.7 | - | 46.3 - 49.1 |
| 2014 UZ224 | 2025 - 2026 | 20.6 | - | - | - | 75.1 - 77.8 |
| 2015 KH162 | 2038 - 2040 | 15.8 | 20.2 | - | - | 68.8 - 71.2 |
| 2015 RR245 | 2025 - 2027 | 17.2 | - | - | - | 46.3 - 51.1 |
| Albion | 2025 - 2030 | 9.7 | 12.3 | 18.5 | - | 42.1 - 43.4 |
| Altjira | 2025 - 2032 | 10.7 | 12.5 | 14.7 | - | 46.6 - 47.0 |
| Borasisi | 2025 - 2027 | 16 | - | - | - | 44.0 - 44.9 |
| Chaos | 2025 - 2034 | 9.6 | 11.1 | 13.5 | - | 41.0 - 41.7 |
| Deucalion | 2038 - 2040 | 10.4 | 11.3 | 14.6 | - | 41.3 - 41.5 |
| Haumea I | 2033 | 23.9 | - | - | - | 45.4 - 45.5 |
| Haumea II | 2037 - 2040 | 11.5 | 13.7 | - | - | 44.1 - 46.8 |
| Huya | 2040 | 16.2 | 16.2 | 16.2 | - | 36.9 - 37.3 |
| Ixion I | 2038 | 19.1 | - | - | - | 30.4 - 30.7 |
| Ixion II | 2040 | 19.2 | 19.2 | 19.2 | - | 30.4 |
| Lempo | 2025 - 2033 | 7.3 | 8.5 | 10.9 | - | 32.1 - 37.2 |
| Logos I | 2031 - 2033 | 13.6 | 13.6 | 17 | 22.4 | 46.4 - 47.6 |
| Logos II | 2036 - 2040 | 10.5 | 14.7 | - | - | 46.4 - 48.2 |
| Makemake I | 2032 - 2033 | 19.2 | 19.3 | 22.5 | - | 52.1 - 52.4 |
| Makemake II | 2036 - 2040 | 12.8 | 19.1 | - | - | 51.6 - 52.5 |
| Manwe | 2025 - 2028 | 14.3 | 22.3 | - | - | 40.0 - 41.2 |
| Mors-Somnus | 2038 - 2040 | 16.7 | 16.7 | 16.7 | - | 41.1 - 42.3 |
| Orcus I | 2027 - 2034 | 10.8 | 12.6 | 15.1 | 20.1 | 45.2 - 47.0 |
| Orcus II | 2036 - 2040 | 13.5 | 23.5 | - | - | 44.1 - 46.3 |
| Otrera | 2025 - 2034 | 6.8 | 7.9 | 10.2 | - | 29.4 - 30.1 |
| Praamzius I | 2028 - 2034 | 9.8 | 11.5 | 13.5 | 17.5 | 43 |
| Praamzius II | 2036 - 2039 | 14.4 | - | - | - | 42.9 - 43.0 |
| Quaoar | 2040 | 20.6 | 20.6 | 20.6 | - | 41.8 |
| Rhadamanthus I | 2032 - 2033 | 20 | 23.7 | 23.8 | - | 44.6 - 44.7 |
| Rhadamanthus II | 2037 - 2040 | 10 | 12.1 | 23.1 | - | 44.2 - 44.9 |
| Salacia | 2025 - 2027 | 15.4 | 24.2 | - | - | 46.1 - 46.4 |
| Sedna | 2025 - 2028 | 18.6 | 22 | - | - | 77.7 - 79.1 |
| Sila-Nunam I | 2028 - 2034 | 9.5 | 11 | 13 | 16.9 | 43.4 |
| Sila-Nunam II | 2036 - 2040 | 13.8 | - | - | - | 43.4 |
| Teharonhiawako | 2025 - 2026 | 19.1 | - | - | - | 45.2 |
| Varda | 2038 - 2040 | 12.6 | 12.6 | 14.6 | - | 40.8 - 42.1 |
| Varuna I | 2025 - 2034 | 10.9 | 12.5 | 14.8 | 20.4 | 44.8 - 45.2 |
| Varuna II | 2036 - 2037 | 19.8 | - | - | - | 45.1 - 45.2 |

**F. Saturn-Uranus Mission Summaries**

Uranus is accessible through a Saturn flyby every year that a Saturn launch is available, between the years 2025-2034 (Uranus I) and 2036-2040 (Uranus II). Because spacecraft arriving from Saturn fly by Uranus earlier than the Jupiter gravity assist missions, the list of KBOs accessible by Saturn-Uranus shares some overlap with the Jupiter-Uranus case. Table 6 lists the possible Saturn-Uranus objects. Comparing the two tables, for some launch vehicles, it is more favorable to fly by Uranus with a Saturn gravity assist than a Jupiter one, such as with (14878) Altjira, (47171) Lempo and (385571) Otrera.

**Table 6: List of KBOs with Saturn and Uranus swingbys, their possible launch years, and the minimum mission duration for various amounts of maximum C3.**

| Object | Launch years | Min Duration C3 < 200 $km^2/s^2$ (years) | Min Duration C3 < 165 $km^2/s^2$ (years) | Min Duration C3 < 140 $km^2/s^2$ (years) | Min Duration C3 < 120 $km^2/s^2$ (years) | Final Target Sun range AU |
|---|---|---|---|---|---|---|
| Uranus I | 2025 - 2034 | 4.4 | 5.2 | 6.3 | 11.5 | 42.9 - 44.8 |
| Uranus II | 2036 - 2040 | 9 | 14.6 | - | - | 44.1 - 44.9 |
| 2002 TC302 | 2026 - 2032 | 18.9 | 21.1 | 22.5 | - | 39.1 - 39.4 |
| 2002 UX25 | 2029 - 2033 | 16.5 | 17.8 | 20.3 | - | 36.7 - 37.0 |
| 2003 AZ84 I | 2027 - 2033 | 17.7 | 17.7 | 17.7 | - | 38.3 - 40.8 |
| 2003 AZ84 II | 2036 - 2040 | 20.8 | - | - | - | 37.0 - 38.3 |
| 2007 UK126 | 2025 - 2034 | 9.2 | 10.3 | 12.8 | - | 37.5 - 38.6 |
| Altjira | 2026 - 2032 | 12.2 | 13.2 | 15.7 | - | 46.7 - 47.0 |
| Chaos | 2025 - 2034 | 9.4 | 10.9 | 13.1 | - | 41.0 - 41.7 |
| Lempo | 2025 - 2033 | 10.7 | 12 | 13.2 | - | 33.1 - 37.2 |
| Otrera | 2025 - 2034 | 11 | 12 | 13.8 | - | 29.6 - 30.1 |
| Praamzius I | 2025 - 2033 | 22.8 | 22.8 | 22.8 | - | 43 |
| Praamzius II | 2038 - 2040 | 22.2 | - | - | - | 42.9-43.0 |
| Sila-Nunam | 2038 - 2040 | 23.1 | - | - | - | 43.4 |
| Varuna I | 2025 - 2029 | 17.6 | 17.6 | 17.6 | - | 44.9 - 45.2 |
| Varuna II | 2036 - 2037 | 22.8 | - | - | - | 45.2 |

### G. Saturn-Neptune Mission Summaries

Neptune is accessible from Saturn between the years 2025-2030. Saturn-Neptune gravitational assists can be used to access a selection of the KBOs accessible via the Jupiter-Neptune path in the early 2030s. Unfortunately, only high C3 launch vehicles could be used, and mission durations do not compare favorably with travel times using a Jupiter gravity assist. Table 7 lists the 8 KBOs accessible via Saturn and Neptune.

**Table 7: List of KBOs with Saturn and Neptune swingbys, their possible launch years, and the minimum mission duration for various amounts of maximum C3.**

| Object | Launch years | Min Duration C3 < 200 $km^2/s^2$ (years) | Min Duration C3 < 165 $km^2/s^2$ (years) | Min Duration C3 < 140 $km^2/s^2$ (years) | Min Duration C3 < 120 $km^2/s^2$ (years) | Final Target Sun range AU |
|---|---|---|---|---|---|---|
| Neptune | 2025 - 2030 | 9.7 | 14.3 | - | - | 29.8 |
| 2003 OP32 | 2025 | 19.8 | - | - | - | 45.2 - 45.4 |
| 2005 QU182 | 2025 | 24.5 | - | - | - | 69.4 |
| 2005 RN43 | 2025 - 2027 | 15.9 | - | - | - | 40.6 - 40.7 |
| 2015 RR245 | 2025 - 2027 | 18 | - | - | - | 46.4 - 50.7 |
| Borasisi | 2025 - 2027 | 16 | - | - | - | 44.0 - 44.9 |
| Manwe | 2025 - 2028 | 14.4 | 22.8 | - | - | 40.0 - 41.2 |
| Salacia | 2025 - 2026 | 19.4 | - | - | - | 46.2 - 46.4 |
| Teharonhiawako | 2025 - 2026 | 18.5 | - | - | - | 45.2 |

**H. Additional Mission Concerns**

Throughout this study we have neglected to consider limits on spacecraft arrival velocity at the target. *New Horizons* successfully imaged Pluto with a velocity of 13.8 km/s [16]. Many of these trajectories have much faster arrival velocities, usually in the 17-19 km/s range, and receiving a second boost from Saturn can put a trajectory into the 25 km/s range. These fast trajectories reduce encounter time, and put the spacecraft images at risk of being blurred, which in turn might require be mitigation by increasing the encounter distance. Slower trajectories, at the cost of TOF, are certainly possible.

Allocating memory to large, but important, spectral and multi-color observations takes up a large amount of close approach time during the flyby [39]. Designing a spacecraft that permits many instruments to work at once, allows memory allocation while other processes are going on, or has smaller memory allocation time will increase the time available for data collection.

While we have kept the arbitrary 25-year limit, we have to consider what is acceptable for a primary mission. *New Horizons*' 9.5-year travel time to its primary mission target was one of the longest in history. While it took *Voyager 2* 12 years to get to Neptune, and *Voyager 2* was built to make it there, its original mission was for five years. Its visit to Uranus and Neptune were extensions of its primary mission [40]. Both *Voyager* spacecraft are still operating (albeit weakly) after 40 years, and the *Cassini* mission ran just one month shy of its 20th launch anniversary. *Opportunity* worked a decade and a half beyond its 90-day life expectancy. A 22-year mission to Eris may be configured as an 8-year mission to Neptune with an Eris extension, even though adding a visit to an ice giant will add a few years to the mission duration.

## I. Selected Mission Recommendations

With the many mission types, KBOs, and possible constraints considered here, what are the best options for a KBO flyby? Table 8 presents the circumstances for a selection of noteworthy missions, including dates of launch, swingby and arrival, C3, mission duration, closest spacecraft approach to body center ($r_q$), plane crossing distance to body center (PC), arrival velocity ($v_\infty$), and KBO heliocentric range at flyby. While the launch and first swingby dates have times of 0 UT for simplicity, the time of day of the second swingby and KBO that were calculated by to minimize ΔV have been truncated. We note that $r_q$ and plane crossing distance are in units of body radii, 71,492 km, 60,268 km, 25,559 km and 24,764 km for Jupiter, Saturn, Uranus, and Neptune respectively[***]. We remind the reader that we have constrained the spacecraft closest approaches to be greater than 6.0, 1.1, 1.1, and 1.1 planetary radii, and the ring plane crossings to be greater than 3.2, 3.0, 2.0, 2.5 planetary radii. For Uranus, spacecraft passage was further restricted to avoid crossing between 2.6-2.7 Uranus radii and between 3.4-4.0 Uranus radii. Many of the mission trajectories fall at the edges of these limits.

Only a selection of trajectories have been displayed in Table 8. We have chosen to highlight the double planet swingby routes that do not incur an excessive time penalty to add a second swingby, compared with only a Jupiter gravity assist. With the exception of Sedna, where only one trajectory was found, we have chosen to highlight 1-2 possible trajectories of what can be up to hundred possible trajectories that met our guidelines met the criteria from

[***] These equatorial radii values are from the pck00010.tpc planetary frames kernel, available for download at https://naif.jpl.nasa.gov/pub/naif/generic_kernels/pck/.

Section II-2. Missions were hand-selected to have a relatively low C3, and a travel time toward the low end of possible mission durations. If a second mission was selected, the focus was on finding a mission with a lower KBO $v_\infty$ while attempting to keep C3 low. Lower KBO $v_\infty$ is often possible for shorter mission durations at the expense of C3. This list is not meant to be exhaustive, but rather it provides a sampling of what is possible, and serves as springboard for further discussion and advocacy.

**Table 8: List of KBOs with Saturn and Neptune swingbys, their possible launch years, and the minimum mission duration for various amounts of maximum C3.**

| KBO name | C3 (km$^2$/s$^2$) | launch date | duration (years) | KBO range (AU) | KBO $v_\infty$ (km/s) | KBO arrival |
|---|---|---|---|---|---|---|
| | **swing 1 name** | **swing 1 date** | **swing 1 $r_q$ (radii)** | **swing 1 PC (radii)** | **swing 1 $v_\infty$ (km/s)** | |
| | **swing 2 name** | **swing 2 date** | **swing 2 $r_q$ (radii)** | **swing 2 PC (radii)** | **swing 2 $v_\infty$ (km/s)** | |
| 2003 AZ84 | 115.8 | 2035-07-23 | 13.7 | 40 | 16 | 2049-03-17 |
| | Jupiter | 2036-11-09 | 21.7 | 24.9 | 12.8 | |
| | Uranus | 2041-07-07 | 1.6 | 2.2 | 13.4 | |
| 2005 QU182 | 134.4 | 2032-04-09 | 15.9 | 68.5 | 18.4 | 2048-02-28 |
| | Jupiter | 2033-05-29 | 6 | 31.3 | 16.1 | |
| | Neptune | 2039-04-15 | 4.4 | 50.6 | 20.7 | |
| 2005 QU182 | 122.1 | 2032-04-04 | 17.8 | 69.5 | 16.2 | 2050-01-07 |
| | Jupiter | 2033-06-18 | 7.6 | 36.8 | 15 | |
| | Neptune | 2039-11-20 | 5.9 | 81.6 | 18.7 | |
| 2005 RN43 | 184 | 2025-05-31 | 19.2 | 40.6 | 8.5 | 2044-08-14 |
| | Saturn | 2027-07-20 | 16.2 | 700.1 | 17.4 | |
| | Neptune | 2036-07-14 | 12.2 | 12.3 | 10.4 | |
| 2007 UK126 | 144.9 | 2035-07-28 | 10.1 | 37.5 | 17.5 | 2045-08-24 |
| | Jupiter | 2036-09-05 | 10.3 | 12.2 | 16.1 | |
| | Uranus | 2040-01-30 | 2 | 2.2 | 19.6 | |
| 2007 UK126 | 98.6 | 2034-06-13 | 12.8 | 37.6 | 13.9 | 2047-04-18 |
| | Jupiter | 2035-11-25 | 6.2 | 6.2 | 10.7 | |
| | Uranus | 2040-04-20 | 2.9 | 3.2 | 16.3 | |
| 2007 UK126 | 156.4 | 2027-06-25 | 10.9 | 38 | 18.3 | 2038-05-11 |
| | Saturn | 2029-11-21 | 3.2 | 3.5 | 14.3 | |
| | Uranus | 2032-12-15 | 2.9 | 2.9 | 17.4 | |
| 2007 UK126 | 169.4 | 2029-07-24 | 11.5 | 37.7 | 15.5 | 2041-01-07 |
| | Saturn | 2031-10-07 | 10.8 | 11.8 | 15.7 | |
| | Uranus | 2034-10-21 | 4.5 | 4.5 | 15.5 | |
| 2013 FY13 | 119 | 2035-07-23 | 22.2 | 74.6 | 18.7 | 2057-09-25 |
| | Jupiter | 2036-10-30 | 19.3 | 22.3 | 13.2 | |
| | Uranus | 2041-03-28 | 1.1 | 14.6 | 14.3 | |
| 2014 EZ51 | 128.6 | 2038-11-04 | 13 | 49.2 | 20.7 | 2051-11-07 |
| | Jupiter | 2040-02-27 | 18.4 | 36.9 | 14.2 | |
| | Saturn | 2041-07-03 | 2.5 | 3 | 14.9 | |
| 2014 EZ51 | 118.6 | 2038-11-09 | 17.9 | 48.2 | 13.9 | 2056-10-09 |
| | Jupiter | 2040-05-17 | 35.6 | 81.8 | 11.2 | |
| | Saturn | 2042-03-25 | 7 | 9 | 10.2 | |
| 2015 KH162 | 128.3 | 2037-10-10 | 17.5 | 68.9 | 19.2 | 2055-04-11 |
| | Jupiter | 2039-02-22 | 6 | 41.1 | 12.6 | |
| | Saturn | 2040-08-15 | 2.8 | 8.1 | 16.5 | |
| 2015 KH162 | 171.1 | 2039-12-14 | 20.6 | 70.1 | 14.8 | 2060-07-14 |
| | Jupiter | 2041-01-22 | 166.7 | 788.8 | 18.1 | |
| | Saturn | 2042-05-08 | 3.9 | 15.4 | 15 | |
| 2015 RR245 | 131.3 | 2032-04-04 | 12.1 | 50.3 | 21.9 | 2044-05-10 |
| | Jupiter | 2033-05-29 | 6.1 | 30.4 | 16.1 | |
| | Neptune | 2039-04-25 | 4.3 | 10.1 | 20.6 | |
| 2015 RR245 | 112.3 | 2032-03-30 | 14.9 | 48.9 | 17.2 | 2047-02-10 |
| | Jupiter | 2033-07-13 | 10.4 | 45.3 | 13.7 | |
| | Neptune | 2040-10-22 | 6.1 | 13.3 | 16.2 | |
| Altjira | 161 | 2027-06-20 | 13.2 | 46.7 | 16.3 | 2040-09-17 |
| | Saturn | 2029-11-01 | 3 | 3.2 | 14.7 | |
| | Uranus | 2032-10-28 | 4.1 | 4.1 | 18 | |
| Altjira | 165 | 2028-07-09 | 13.9 | 46.7 | 14.8 | 2042-06-13 |
| | Saturn | 2030-10-17 | 5.1 | 5.5 | 15.2 | |
| | Uranus | 2033-09-22 | 4.9 | 4.9 | 16.9 | |
| Borasisi | 134.4 | 2032-04-09 | 11 | 44.2 | 18.9 | 2043-04-07 |
| | Jupiter | 2033-05-29 | 6 | 31.3 | 16.1 | |
| | Neptune | 2039-04-15 | 1.7 | 2.8 | 20.7 | |
| Borasisi | 119 | 2032-04-09 | 13.5 | 44.4 | 14.9 | 2045-09-25 |
| | Jupiter | 2033-07-08 | 9.6 | 46.1 | 14.1 | |
| | Neptune | 2040-07-15 | 2.7 | 4.2 | 16.9 | |
| Chaos | 157.7 | 2035-07-28 | 9.7 | 41.1 | 19.6 | 2045-04-18 |
| | Jupiter | 2036-08-16 | 8.2 | 9.8 | 17.3 | |
| | Uranus | 2039-10-06 | 2.1 | 2.1 | 21.5 | |
| Chaos | 135.8 | 2035-07-23 | 11.7 | 41.2 | 15.7 | 2047-04-13 |
| | Jupiter | 2036-09-20 | 12.3 | 14.4 | 15.2 | |
| | Uranus | 2040-05-12 | 2.9 | 2.9 | 18 | |
| Chaos | 163.2 | 2027-06-25 | 10.9 | 41 | 19.5 | 2038-06-04 |
| | Saturn | 2029-10-17 | 2.8 | 3 | 15.1 | |
| | Uranus | 2032-09-19 | 2.6 | 3 | 18.5 | |
| Chaos | 166.5 | 2029-07-19 | 12.5 | 41.1 | 15 | 2042-02-01 |
| | Saturn | 2031-10-17 | 11.4 | 12.4 | 15.4 | |
| | Uranus | 2034-11-27 | 3.8 | 4.1 | 15.1 | |
| Deucalion | 154.4 | 2039-11-24 | 11.9 | 41.5 | 17.5 | 2051-10-10 |
| | Jupiter | 2041-02-06 | 187 | 738.3 | 16.9 | |
| | Saturn | 2042-07-16 | 2.6 | 3 | 13.4 | |
| Deucalion | 119.5 | 2038-10-30 | 11.9 | 41.5 | 18.7 | 2050-09-15 |
| | Jupiter | 2040-03-18 | 22 | 589.1 | 13.3 | |
| | Saturn | 2041-09-01 | 2.8 | 3.1 | 13.6 | |
| Eris | 134.4 | 2032-04-09 | 22.3 | 92.3 | 19.9 | 2054-07-26 |
| | Jupiter | 2033-05-29 | 6 | 31.3 | 16.1 | |
| | Neptune | 2039-04-15 | 5.7 | 6.4 | 20.7 | |
| Eris | 122.1 | 2032-04-04 | 24.9 | 91.9 | 17.6 | 2057-02-13 |
| | Jupiter | 2033-06-18 | 7.6 | 36.8 | 15 | |
| | Neptune | 2039-11-20 | 6.3 | 8.8 | 18.7 | |
| Haumea | 128.3 | 2037-10-10 | 12.8 | 46.7 | 19.2 | 2050-07-22 |
| | Jupiter | 2039-02-22 | 6 | 41.1 | 12.6 | |
| | Saturn | 2040-08-15 | 4.9 | 21.1 | 16.5 | |
| Haumea | 124.3 | 2038-10-30 | 14.9 | 46.2 | 15 | 2053-09-20 |
| | Jupiter | 2040-03-03 | 19.3 | 614.3 | 13.9 | |
| | Saturn | 2041-07-19 | 8 | 39.5 | 14.5 | |
| Lempo | 169.4 | 2029-07-24 | 11.8 | 33.4 | 11.5 | 2041-05-11 |
| | Saturn | 2031-10-07 | 10.8 | 11.8 | 15.7 | |
| | Uranus | 2034-10-21 | 2.3 | 2.3 | 15.5 | |
| Logos | 128.3 | 2037-10-10 | 13.4 | 46.8 | 15.8 | 2051-03-18 |
| | Jupiter | 2039-02-22 | 6 | 41.1 | 12.6 | |
| | Saturn | 2040-08-15 | 41.1 | 92.2 | 16.5 | |
| Makemake | 120.2 | 2037-10-05 | 16.9 | 52.3 | 14.2 | 2054-09-04 |
| | Jupiter | 2039-02-22 | 6.1 | 39.4 | 12.6 | |
| | Saturn | 2040-08-17 | 7.2 | 38.2 | 16.5 | |
| Manwe | 134.4 | 2032-04-09 | 10 | 41 | 20.1 | 2042-03-24 |
| | Jupiter | 2033-05-29 | 6 | 31.3 | 16.1 | |
| | Neptune | 2039-04-15 | 2.2 | 3.6 | 20.7 | |
| Manwe | 164.6 | 2033-05-14 | 11.8 | 40.7 | 15.5 | 2045-02-11 |
| | Jupiter | 2034-05-14 | 25.4 | 68.3 | 18.8 | |
| | Neptune | 2041-03-28 | 4 | 6.5 | 16.3 | |
| Manwe | 178.1 | 2025-05-31 | 18.3 | 40.9 | 9.1 | 2043-09-09 |
| | Saturn | 2027-08-09 | 17.9 | 852.7 | 16.8 | |
| | Neptune | 2037-04-17 | 51.3 | 72 | 9.6 | |
| Otrera | 158.3 | 2029-07-19 | 12.4 | 29.6 | 10.4 | 2041-12-18 |
| | Saturn | 2031-11-21 | 13.6 | 14.8 | 14.5 | |
| | Uranus | 2035-03-31 | 2.2 | 2.2 | 13.8 | |
| Rhadamanthus | 120.2 | 2037-10-05 | 11.5 | 44.3 | 20.1 | 2049-03-18 |
| | Jupiter | 2039-02-22 | 6.1 | 39.4 | 12.6 | |
| | Saturn | 2040-08-17 | 5 | 7.5 | 16.5 | |
| Rhadamanthus | 125.5 | 2038-11-04 | 13.2 | 44.5 | 15.8 | 2052-01-04 |
| | Jupiter | 2040-03-08 | 20 | 41 | 13.8 | |
| | Saturn | 2041-08-01 | 7.8 | 13 | 14.3 | |
| Sedna | 134.4 | 2032-04-19 | 24.9 | 77.2 | 17.5 | 2057-03-08 |
| | Jupiter | 2033-08-02 | 12.7 | 62.4 | 13 | |
| | Neptune | 2041-07-28 | 1.1 | 8.5 | 14.5 | |
| Varda | 115.7 | 2038-11-04 | 14.9 | 42.1 | 16.3 | 2053-09-26 |
| | Jupiter | 2040-04-17 | 28.1 | 545.2 | 12.2 | |
| | Saturn | 2041-12-05 | 1.1 | 3.2 | 11.9 | |
| Varuna | 128.5 | 2035-07-28 | 12.9 | 45 | 17.8 | 2048-06-03 |
| | Jupiter | 2036-10-10 | 15.2 | 17.8 | 14.2 | |
| | Uranus | 2040-09-27 | 1.4 | 2.2 | 16.3 | |
| Varuna | 117.8 | 2035-07-18 | 15 | 45 | 14.8 | 2050-06-29 |
| | Jupiter | 2036-11-04 | 20.7 | 23.5 | 12.9 | |
| | Uranus | 2041-05-27 | 2.1 | 3.2 | 13.7 | |
| Varuna | 98.6 | 2034-06-13 | 13.5 | 45 | 18 | 2047-12-28 |
| | Jupiter | 2035-11-25 | 6.2 | 6.2 | 10.7 | |
| | Uranus | 2040-04-20 | 1.5 | 2.3 | 16.3 | |

## IV. Future Work and Conclusions

We have presented the launch years and minimum travel times for Pluto and 45 of the most notable KBOs-- all named objects and all objects with an H magnitude greater than 4. This is not an exhaustive list of mission objects, though it does cover the most desirable and scientifically interesting targets.

There is a KBO mission possible for every Earth-Jupiter launch window throughout a Jupiter revolution, thus Pluto and every one of the selected 45 KBOs are accessible via Jupiter gravity assist with a flight time of under 25 years and a C3 less than 140 $km^2/s^2$. Many, but not all objects can be reached via Saturn flyby, and a smaller list still can be compatible with a visit to an ice giant, though it does not necessarily provide a TOF advantage.

There is a quadrant of the solar system with launch windows for a Jupiter gravity assist in the late 2020s that do not share opportunities with any other giant planet. This group includes (225088) 2007 OR10, 2010 RF43, (15810) Arawn, and Pluto. These objects, if they are to be visited, must have dedicated missions in our time period, or wait until Saturn swings into view in the 2040s.

We found that all five of the non-Pluto KBOs studied by McGranaghan *et al* [23] can be reached by giant planet swingby—(136199) Eris and (90377) Sedna with Neptune, and (50000) Quaoar, (136472) Makemake and (136108) Haumea via Jupiter-Saturn. Fast-rotator (20000) Varuna is reachable after a Uranus encounter.

While we have greatly expanded the number of KBOs that have had mission searches performed, and we have included ice giants in our search, we have not exhausted the types of KBO missions that are workable, such as direct Uranus/Neptune launches, or Venus/Earth gravity assists, considered by Guo et al. for *New Horizons*, and Hansen for *Argo*. Further prospects might be found if longer travel times to Saturn are considered.

We have not considered multiple KBO flybys. It is worth noting that the prospect of an additional flyby of a KBO after Pluto played a large role in *New Horizons*' selection, and that it was possible because of Pluto's proximity at the time to the plane of the Solar System. Many of our KBOs have highly inclined orbits and are often far outside the plane of the Solar System at the time of flyby, thus no additional flybys will be possible en route owing to the vast separations between objects. To probe a second KBO typically requires one to fly through the classical belt.

We have also neglected to examine any Centaur missions in this paper.

If a mission to the Kuiper Belt via Saturn, Uranus or Neptune is to occur, the next launch opportunities are in the 2030s: first to Neptune, then to Uranus, then to Saturn. Saturn-Uranus missions would launch in the late 2020s. These missions must be called for in the next decadal survey, and more concrete studies should be undertaken immediately, or the opportunity for them will soon be lost.

## Funding Sources

This project was funded in its entirety by internal research funding from Southwest Research Institute under Presidential Discretion IR&D, Proposal 15-80846 "Pluto-Kuiper Belt Mission Studies."

## Acknowledgments

The authors would like to thank Simon B. Porter, and Emma Birath for their help with some initial MAKO coding work. We acknowledge the helpful comments from two anonymous reviewers.